\documentclass[%
preprint,
superscriptaddress,
 amsmath,amssymb,
 aps,
 prl,
aps,prl]{revtex4-2}

\usepackage{graphicx}
\usepackage{dcolumn}
\usepackage{bm}
\usepackage{float}

\begin{document}

\title{Phase-Dependent Squeezing in Dual-Comb Interferometry}

\author{Daniel I. Herman}
\email{daniel.i.herman@colorado.edu}
\email{Present Address: Sandia National Laboratories, Albuquerque, NM}
\affiliation{Department of Electrical, Computer and Energy Engineering, University of Colorado Boulder, Boulder, CO 80309, USA}
\author{Molly Kate Kreider}
\affiliation{Department of Electrical, Computer and Energy Engineering, University of Colorado Boulder, Boulder, CO 80309, USA}
\affiliation{Department of Physics, University of Colorado Boulder, Boulder, CO 80309, USA}
\author{Noah Lordi}
\affiliation{Department of Physics, University of Colorado Boulder, Boulder, CO 80309, USA}
\author{Mathieu Walsh}
\affiliation{Centre d’Optique, Photonique et Laser, Université Laval, Québec, Québec G1V 0A6, Canada}
\author{Eugene J. Tsao}
\affiliation{Department of Electrical, Computer and Energy Engineering, University of Colorado Boulder, Boulder, CO 80309, USA}
\author{Alexander J. Lind}
\affiliation{Department of Electrical, Computer and Energy Engineering, University of Colorado Boulder, Boulder, CO 80309, USA}
\author{Matthew Heyrich}
\affiliation{Department of Electrical, Computer and Energy Engineering, University of Colorado Boulder, Boulder, CO 80309, USA}
\affiliation{Department of Physics, University of Colorado Boulder, Boulder, CO 80309, USA}
\author{Joshua Combes}
\affiliation{Department of Electrical, Computer and Energy Engineering, University of Colorado Boulder, Boulder, CO 80309, USA}
\author{Scott A. Diddams}
\email{scott.diddams@colorado.edu}
\affiliation{Department of Electrical, Computer and Energy Engineering, University of Colorado Boulder, Boulder, CO 80309, USA}
\affiliation{Department of Physics, University of Colorado Boulder, Boulder, CO 80309, USA}
\author{Jérôme Genest}
\email{jgenest@gel.ulaval.ca}
\affiliation{Centre d’Optique, Photonique et Laser, Université Laval, Québec, Québec G1V 0A6, Canada}

\date{\today}

\begin{abstract}
We measure phase-dependent Kerr soliton squeezing and anti-squeezing in the time-domain dual-comb interferograms generated using two independent frequency comb lasers. The signal appears as non-stationary quantum noise that varies with the fringe phase of the interferogram and dips below the shot-noise level by as much as 3.8 dB for alternating zero-crossings. The behavior arises from the periodic displacement of the Kerr squeezed comb by the coherent field of the second frequency comb, and is confirmed by a quantum noise model. These experiments support a route towards quantum-enhanced dual-comb timing applications and raise the prospect of high-speed quantum state tomography with dual-comb interferometry.
\end{abstract}

\maketitle

Optical interferometry with frequency combs is central to their use in fundamental time and frequency metrology at the highest levels of precision. Within this broad landscape, interferometry between pairs of frequency combs, now termed dual-comb interferometry (DCI), is integral to applications that include distance ranging \cite{Coddington09}, free-space optical time transfer \cite{Giorgetta13,Caldwell23}, holography \cite{Vicentini21}, micro-imaging \cite{Ideguchi13,Chang23}, fiber-based optical networking \cite{Chen24}, and spectroscopy \cite{Coddington16}. Frequency comb metrology reaches beyond traditional continuous-wave (CW) laser metrology by harnessing temporal coherence and broad optical bandwidth simultaneously through the full stabilization of the carrier phase and envelope timing of femtosecond optical pulses \cite{Diddams00}. \\
\indent Squeezing of CW laser light has led to quantum noise reduction for both interferometric and spectroscopic measurements, leading to improved gravitational wave detectors and frequency modulation spectroscopy \cite{Ganapathy23,Polzik92}. Theoretical proposals \cite{Lamine08,Belsley23,Shi23} and some initial experiments \cite{Wang18} have also shown that squeezed optical frequency combs can provide quantum metrological advantages. In this context, DCI aids metrology by down-converting the optical comb spectrum to accessible and traceable radio frequency (RF) signals. A DCI experiment consists of two mutually coherent comb sources with slightly different repetition rates ($f_\mathrm{rep,1}$ and $f_\mathrm{rep,2}$ = $f_\mathrm{rep,1} + \Delta f_\mathrm{rep}$) that interfere on a photodiode to yield an interferogram (IGM) signal that is periodic at $\Delta f_\mathrm{rep}$. For spectroscopy, the IGM is Fourier-transformed to recover an RF comb whose amplitude is proportional to the geometric mean of the individual comb spectra \cite{Coddington16}. In time transfer and ranging, the IGM centerburst is cross-correlated and phase-demodulated to yield timing information \cite{Deschenes16}. Recently, frequency combs have reached the quantum limits associated with coherent states for dual-comb timing and ranging experiments \cite{Caldwell22,Caldwell23}.\\
\indent 
In this work, we discuss a squeezed DCI measurement scheme and its potential impact on dual-comb time transfer, ranging and networking by revealing sub-shot-noise-limited statistics on the fringes of a dual-comb IGM. This new form of time-domain pulsed squeezing measurement can also be applied to characterize the quantum states generated by optical nonlinearities \cite{Rasputnyi24}. And while nonlinear electro-optical sampling has been used to characterize ultra-fast quantum states \cite{Riek15,Riek17}, our results show that linear-optical sampling with dual-combs can extract similar information.\\
\begin{figure}[hb]
\includegraphics[width=0.8\linewidth]{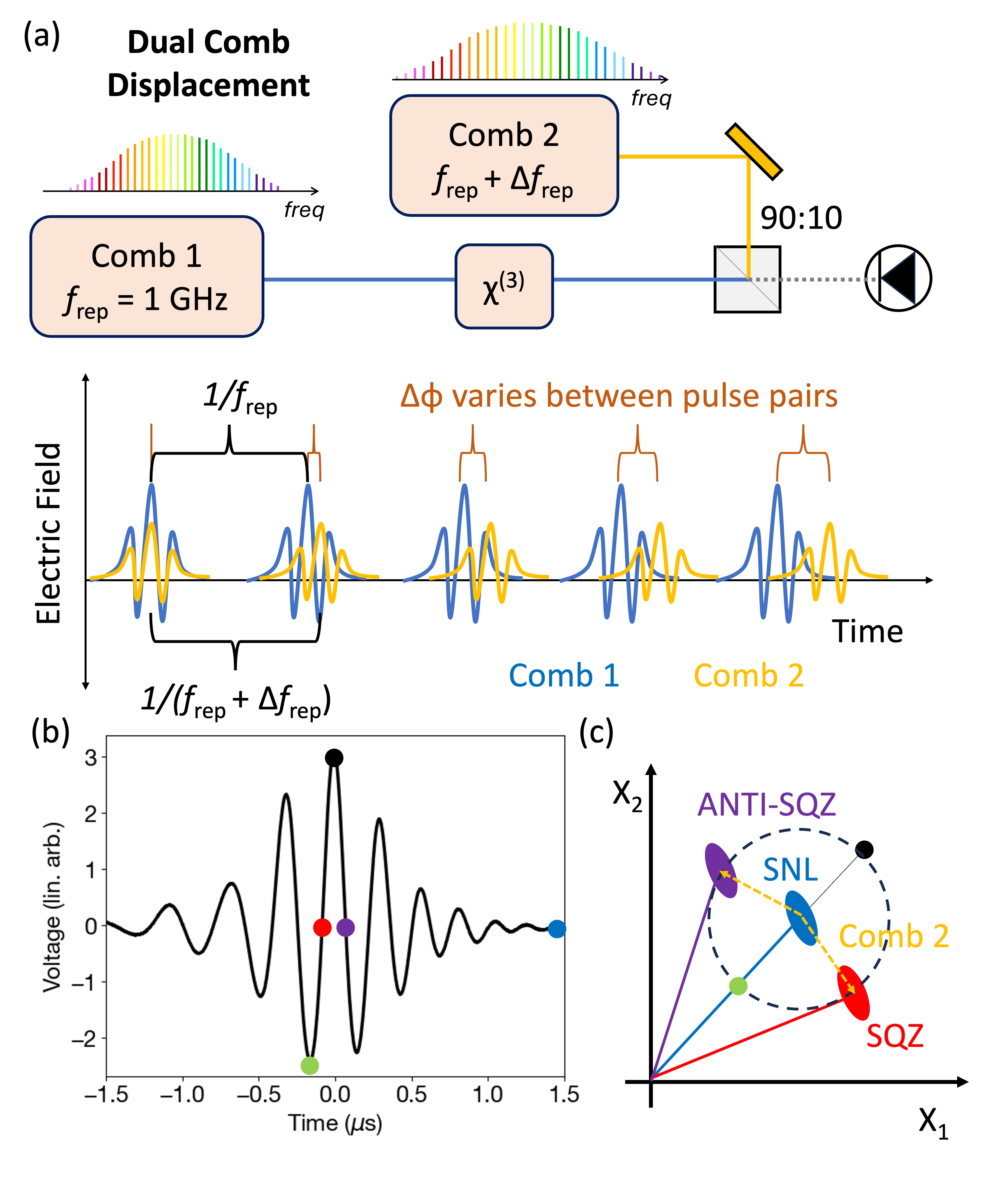}
\caption{ Conceptual schematics. (a) Simplified experimental block diagram and time-domain picture of quantum dual-comb displacement operation utilizing $\chi^{(3)}$ squeezing.  (b) Measured single-shot IGM with several highlighted points: $-\pi/2$ radians (red), $0$ radians (black), $\pi/2$ radians (purple), $\pi$ radians (green) and large path difference (blue). (c) Phase space representation indicating anti-squeezed state (ANTI-SQZ; purple), squeezed state (SQZ; red), IGM maximum (black), IGM minimum (green) and the shot-noise-limited Kerr state at large path differences (SNL; blue).}
\label{fig:1}
\end{figure}
\indent Squeezing of pulsed light was achieved only two years after the first demonstration of CW laser squeezing \cite{Slusher85,Wu86,Slusher87}. Early experiments showed that the quantum noise of pulsed lasers is non-stationary and noise gating reveals pulsed squeezed states with better contrast \cite{Aytur90}. Similar techniques have allowed for improved classical noise limits for frequency comb measurements including microwave generation \cite{Quinlan13}, CW-comb heterodyne comparison \cite{Deschenes13} and dual-comb spectroscopy \cite{Walsh23}. A quantum mechanical formalism has been developed to more generally describe the role of temporal mode matching in dual-comb measurements \cite{Lordi24}. This treatment separates DCI into a quadrature-like measurement regime for small optical path differences near the ``centerburst" region of the IGM and an intensity-like measurement regime for large optical path differences. Recently, we utilized amplitude-squeezing to gain a $\sim$3 dB metrological advantage in dual-comb spectroscopy by reducing quantum noise for the intensity-like measurement regime at large optical path differences \cite{Herman24}. Here, we demonstrate phase-dependent squeezing in the centerburst of a dual-comb signal using a method that resembles a quadrature measurement [see Figs.\thinspace 1(a) and 1(b)].\\
\indent We generate a squeezed comb using the $\chi^{(3)}$ Kerr nonlinearity in a polarization-maintaining highly nonlinear fiber (PM-HNLF). Because the Kerr effect is number-preserving, the squeezing angle of an un-displaced Kerr state yields the same amplitude noise as the original coherent state before Kerr interaction \cite{Kalinin23}. The squeezed quadrature of the Kerr state is revealed through a small optical displacement [see Fig.\thinspace 1(c)] followed by intensity detection \cite{Fiorentino01}. For shot-noise-limited laser sources, squeezing can be measured with a single photodiode \cite{Herman24}. Here, we displace a squeezed pulse using a field derived from an entirely independent laser system. Although advanced squeezing systems (e.g., Laser Interferometer Gravitational Observatory) have demonstrated squeezing-enhanced CW interferometry using two separately generated laser fields \cite{Ganapathy23}, these measurements are performed at low Fourier frequencies where the phase coherence of two sources is established using conventional phase-locking techniques. Our work demonstrates that a quantum displacement of one laser by another can occur for high Fourier frequencies above the bandwidth limit of the phase-locked loops used to establish mutual coherence between the sources.\\
\begin{figure}[t]
\centering
\includegraphics[width=0.8\linewidth]{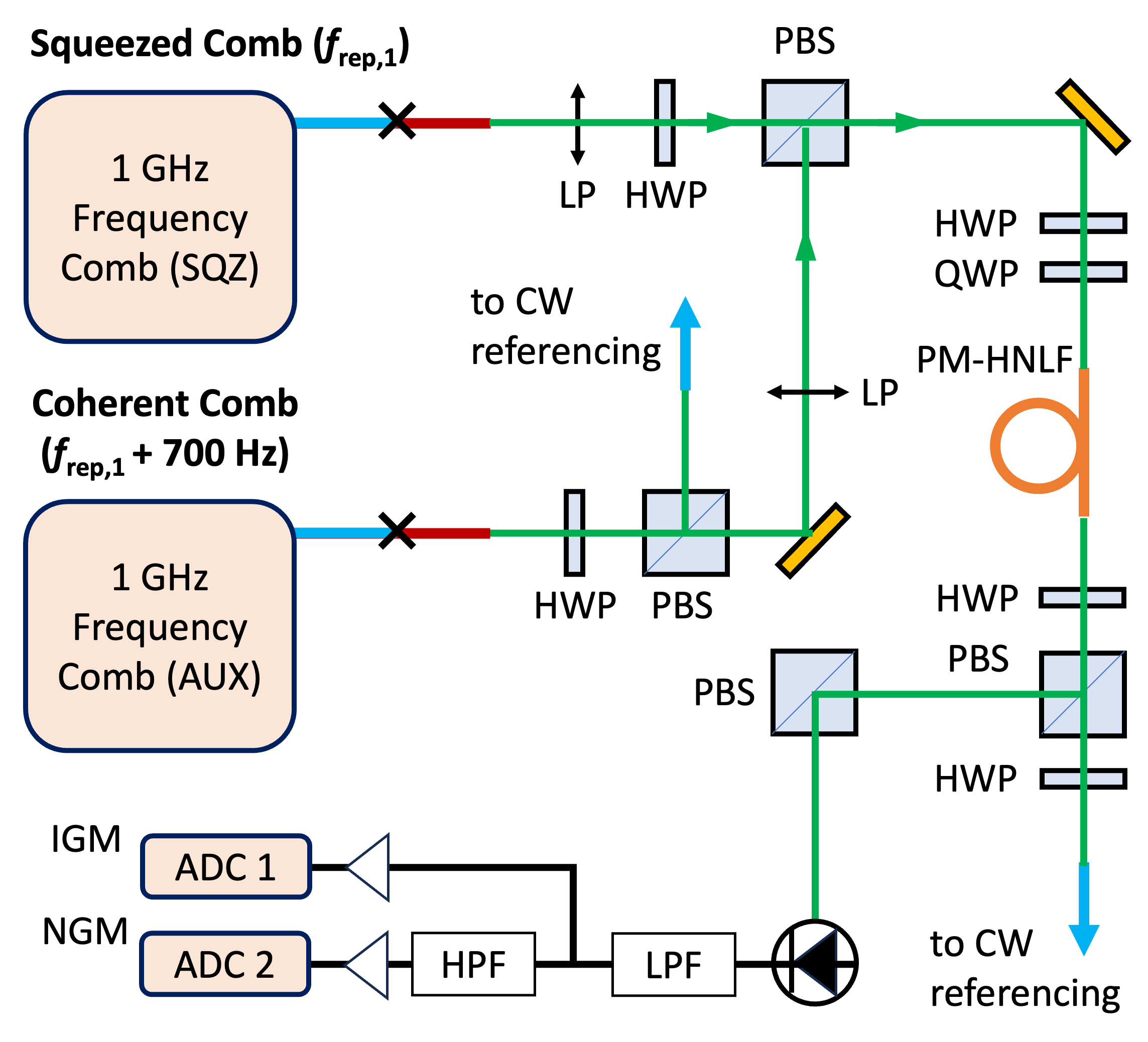}
\caption{Experimental schematic.\ Two 1 GHz combs are optically-isolated and compressed to the transform limit.\ Polarization optics combine the combs with a variable splitting ratio and send cross-polarized combs onto orthogonal axes of a PM-HNLF. The two combs are recombined on a half-wave plate (HWP) and polarization beam splitter (PBS). The IGM is recorded on a photodiode and digitized using a a high-speed analog-to-digital converter (ADC). Both combs are sampled for comparison and locking against a narrow linewidth CW laser. LP: linear polarizer; QWP: quarter-wave plate; LPF: low-pass RF filter; HPF: high-pass RF filter; Blue/red/orange lines indicate PM-1550/PM-ND/PM-HNLF fiber; Green lines indicate free space optical path; Black x's indicate fiber splice; Black lines indicate RF signal.}
\label{fig:2}
\end{figure}
\begin{figure}[htbp!]
\centering
\includegraphics[width=0.45\linewidth]{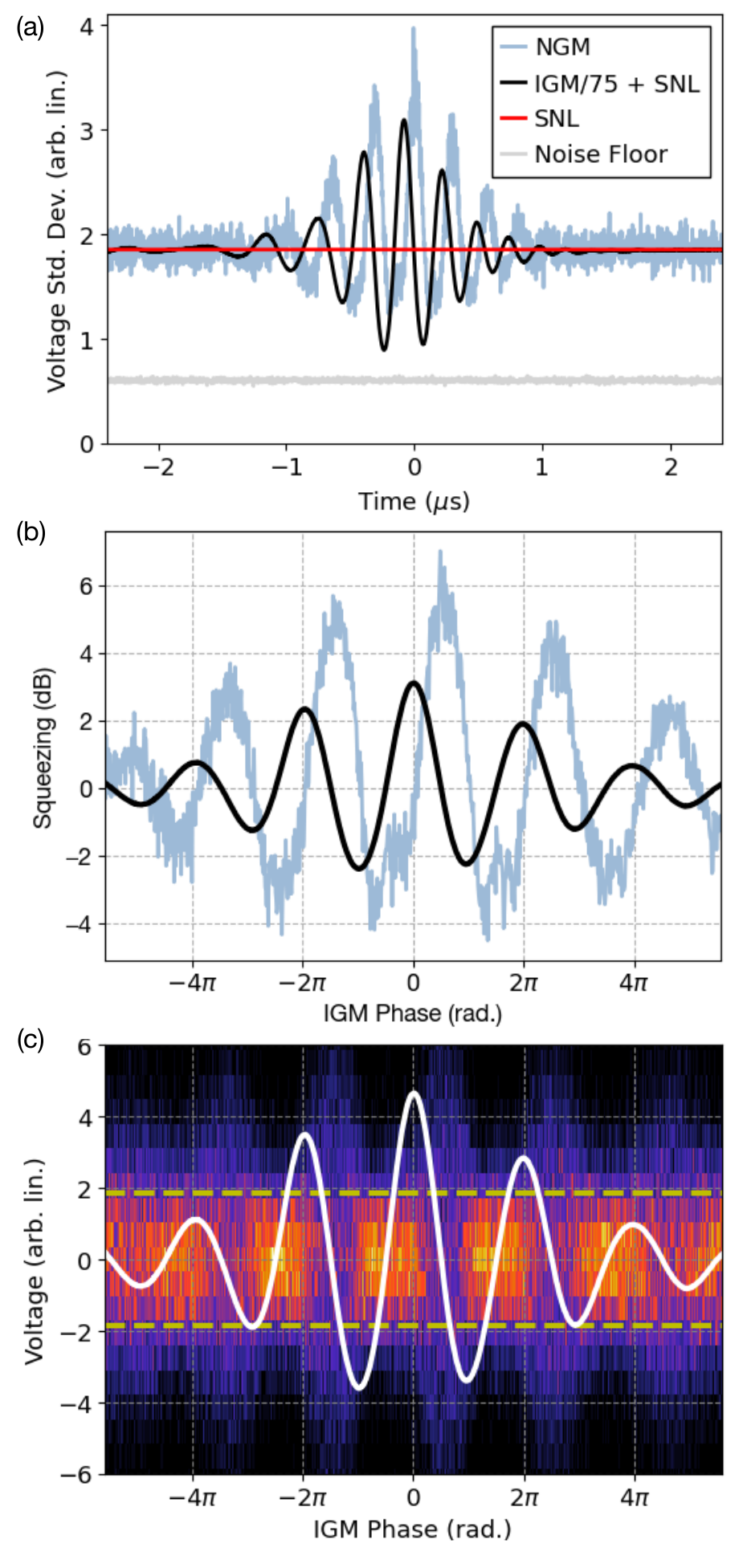}
\caption{Time-dependent quantum noise fluctuations during a dual-comb IGM. (a) Quantum noise reduction and amplification are visible during the centerburst. Blue = noise-gram (NGM); Black = scaled interferogram (IGM); Red = shot-noise limit (SNL); Grey = electrical noise floor. The NGM, SNL and noise floor are plotted as voltage standard deviation, while the IGM is plotted as a scaled voltage. (b) Zoomed-in image of centerburst region showing the squeezing factor $V_{\mathrm{SQZ}}$ plotted in dB relative to the SNL. (c) False color histogram of NGM voltage counts versus unwrapped IGM phase. Brighter colors indicate higher voltage counts. Dotted yellow line = SNL; White line = vertically-scaled IGM.}
\label{fig:3}
\end{figure}
\indent To accomplish the dual-comb displacement operation, we inject two cross-polarized frequency comb lasers into a PM-HNLF (see Fig. 2) at a power ratio of 10:1. Both combs have repetition rates near 1 GHz, are centered at 1563$\pm$1 nm, and are temporally compressed in fiber to $\sim$260 fs. This geometry minimizes non-common technical noise with the stronger comb experiencing significant Kerr squeezing, while the weaker comb remains approximately in a coherent state. Both combs are phase-locked to a narrow linewidth 1564 nm CW laser with feedback to cavity piezo-electric actuators with less than 100 kHz bandwidth. The repetition rate difference of the comb lasers is set to $\sim$700 Hz. After PM-HNLF, the combs are recombined with a power ratio of 100:1. \\
\indent A single high quantum efficiency InGaAs photodiode is used to detect the dual-comb IGMs which are centered at 3 MHz and span 1 MHz to 5 MHz in the RF domain. The IGMs are low-pass filtered at 225 MHz and then split into two signals. The first signal is amplified using a 20 dB gain RF amplifier and recorded using a 500 MS/s analog-to-digital converter (ADC). The second signal is high-pass filtered at 25 MHz to remove the IGM itself and the high frequency noise is then amplified using a low-noise RF amplifier and recorded on a second ADC channel. We refer to the second signal as the ``noise-gram" (NGM). The low IGM center frequency allows us to over-sample the IGM fringe and fully capture the minimum values of the quantum noise. Because the carrier-envelope offset (CEO) frequencies of our combs are not stabilized, we post-select only the IGMs that have similar CEO phases.\\
\begin{figure}[!t]
\centering
\includegraphics[width=0.8\linewidth]{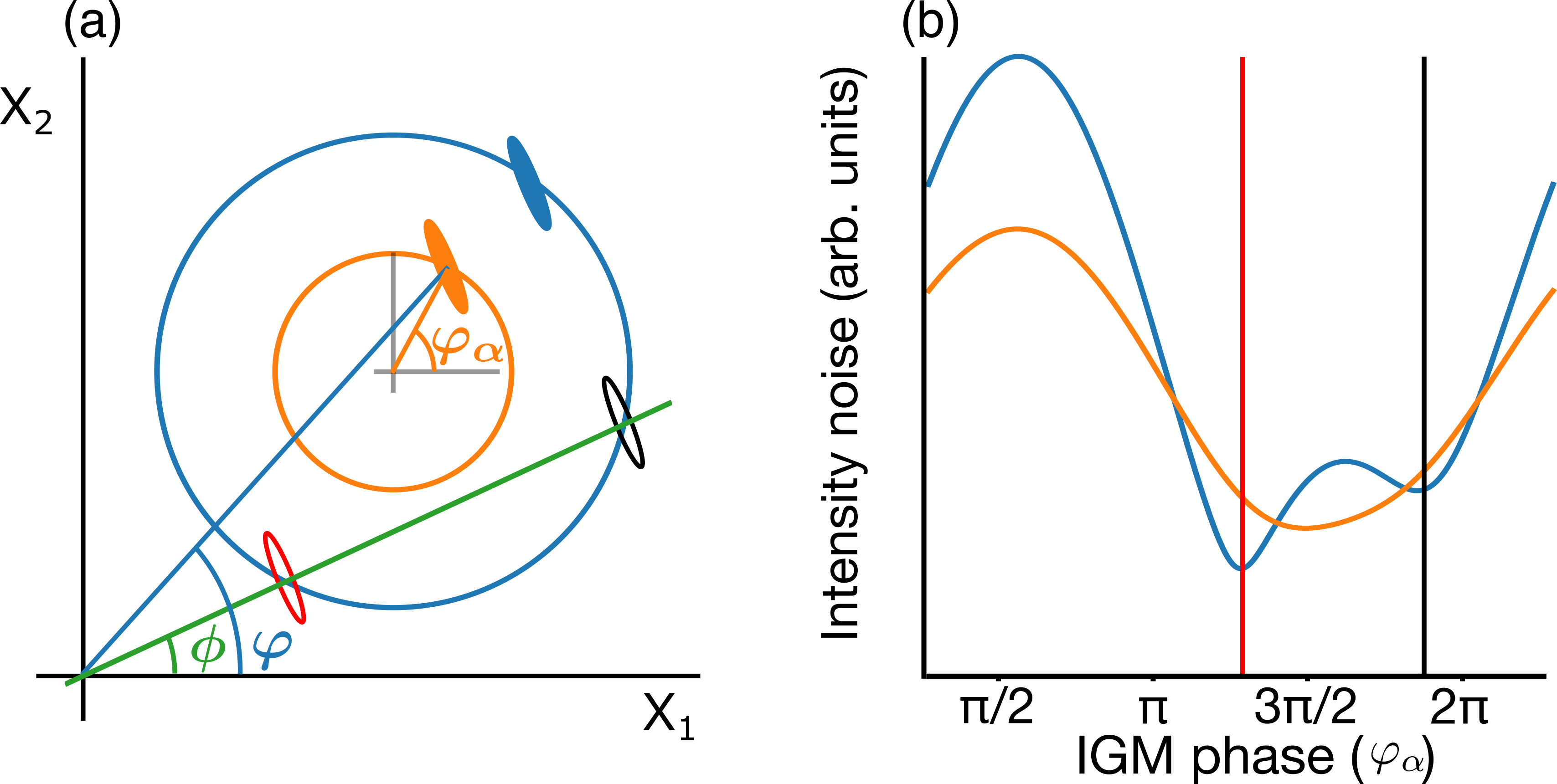}
\caption{Geometric picture of power-dependent noise properties in dual-comb displacement operation. (a) Phase space (Wigner representation) schematic demonstrating the time-varying noise within the dual-comb IGM. The orange phase space path has intensity noise with one local maximum and one local minimum. The larger blue phase space path has intensity noise with two local minima/maxima. The green line indicates the squeezing angle and sets the phase angle for local minima of the amplitude variance. (b) Intensity noise of phase space paths near the bifurcation point. Red/black vertical dotted lines indicate the local minima points that are also marked in (a).}
\label{fig:4}
\end{figure}
\indent A continuous stream of 730 IGMs and NGMs are acquired on two ADC channels. We perform a Hilbert transform on the individual IGMs and then cross-correlate each IGM against the first IGM to align them based on their envelopes. The delays calculated from the IGM cross-correlation are also applied to the NGMs. We then down-select the IGMs and NGMs by the CEO phase calculated using Hilbert transformation. This selection process yields 50 IGMs within a $\Delta_\mathrm{ceo} = f_\mathrm{ceo,1} - f_\mathrm{ceo,2}$ range of $<$50 mrad. After CEO binning, we take the standard deviation along the IGM number axis, \textit{i.e.}, the standard deviation of points from every IGM at the properly aligned time [see Fig. 3(a)]. The high frequency NGM is known to be dominated by shot noise at large path differences \cite{Herman24}, and this noise is well above the electrical noise floor, which allows us to derive the measurement shot noise limit (SNL) from these data. Zooming in on the IGM centerburst region, the NGM exhibits non-stationarity that is phase-coherent with the IGM.\\
\indent We quantify the squeezing factor ($V_\mathrm{SQZ}$) as
\begin{equation}
V_\mathrm{SQZ} = \frac{\sigma^2_{\varphi\alpha}- \sigma^2_\mathrm{NF}}{\sigma^2_\mathrm{SNL}-\sigma^2_\mathrm{NF}},
\end{equation}
where $\sigma^2_{\varphi\alpha}$ is the IGM variance at a given IGM phase $\varphi_\alpha$, $\sigma^2_\mathrm{SNL}$ is the SNL variance at large path differences, and $\sigma^2_\mathrm{NF}$ is the variance of the electrical noise floor. The squeezing results shown in Fig. 3(b) are plotted versus the demodulated, unwrapped and averaged IGM phase axis. The maximum measured squeezing level is $\sim$3.8 dB which is close to the level measured using single-comb self-displacement \cite{Herman24}. We observe a maximum anti-squeezing of $\sim$6 dB. In Fig. 3(c), we plot the full histogram of NGM values for each IGM phase, clearly showing the changing statistics as the IGM phase evolves. Of particular note is the appearance of squeezing and anti-squeezing at alternating zero-crossings of the IGM. This form of non-stationarity should only be present if the displaced state is squeezed. A small amount of non-stationarity is to be expected during an IGM as the power on the detector changes slightly, but for our comb power ratio (14.6 mW Comb 1 and 170 $\mu$W from Comb 2) this effect is small.  For an un-balanced measurement, the varying SNL is expected to be directly correlated with the IGM level rather than shifted to the zero-crossings.\\
\indent The observed statistics can be explained using a simple model of the displacement operation that incorporates the pulse width normalized to the duty cycle of the lasers, the measured squeezing level from the single comb experiment and the evolving mode-mismatch across the IGM (see Supplemental Materials). Specifically, the model predicts that the NGM voltage variance, $V = \sigma^2_{\varphi\alpha}$, takes the following form:
\begin{align}
V &= |\eta|^2|\beta+\gamma\alpha|^2\left(2\sinh^2(r)
-\sinh(2r)\cos(2(\phi-\varphi))\right)\notag\\
        &+ |\eta|\left(|\beta|^2 +|\alpha|^2 + 2\text{Re}(\beta^*\gamma\alpha)\right),
\end{align}
where $\eta$ is detection efficiency (including losses), $\alpha$ is the weak coherent state amplitude (displacing state), $\beta$ is the strong squeezed state amplitude, $r$ is the squeezing parameter, $\gamma$ is the mode-overlap of the pulses, $\phi$ is the squeezing angle and $\varphi$ is phase angle of the mean field state. The mean field phase can be written in terms of the phase of the $\alpha$ state which also acts as the IGM phase ($\varphi_\alpha$) for a frame that co-rotates with the $\beta$ state:
\begin{equation}
\varphi\approx \frac{|\gamma\alpha|\mathrm{sin}(\varphi_\alpha)}{|\beta|}.
\end{equation}
Equation 2 predicts a ``bifurcation point" in the NGMs as the ratio $R = |\gamma\alpha|/|\beta|$ is varied, which is viewed geometrically in Fig. 4(a). For small $R$, the intensity noise-gram has two extrema per 2$\pi$ of displacement phase [see Fig. 4(b)]. This statement is confirmed by Taylor expanding the $\varphi$-dependent term in Eq. (2) for small $R$:
\begin{equation}
\mathrm{cos}(2\phi - 2R\ \mathrm{sin}(\varphi_\alpha)) \approx \mathrm{cos}(2\phi) + 2R\ \mathrm{sin}(2\phi)\mathrm{sin}(\varphi_\alpha).    
\end{equation}
As $R$ increases, two more extrema will appear (four total extrema per IGM cycle) resulting in frequency doubling of the NGM.\\
\indent To confirm the model prediction we conducted a series of measurements using a dual-detector setup where we performed a photocurrent subtraction/addition in post-processing over a range of power ratios. The setup follows Fig. 2 except an additional 50:50 beamsplitter is placed before the photodetector. Two identical high quantum efficiency photodetection chains record the signals at each of the beamsplitter outputs. As before, the signals are separated into IGM/NGM filtered components. A four-channel ADC is used to record the two IGMs and two NGMs. The detection chains use matched cable lengths and components to minimize relative delays which is confirmed by cross-correlating the IGM recorded on each channel. After alignment, summing the two NGM channels yields the NGM measured from a single-channel measurement and differencing the two NGM channels yields the fringe-varying SNL.
\begin{figure}[!b]
\centering
\includegraphics[width=0.8\linewidth]{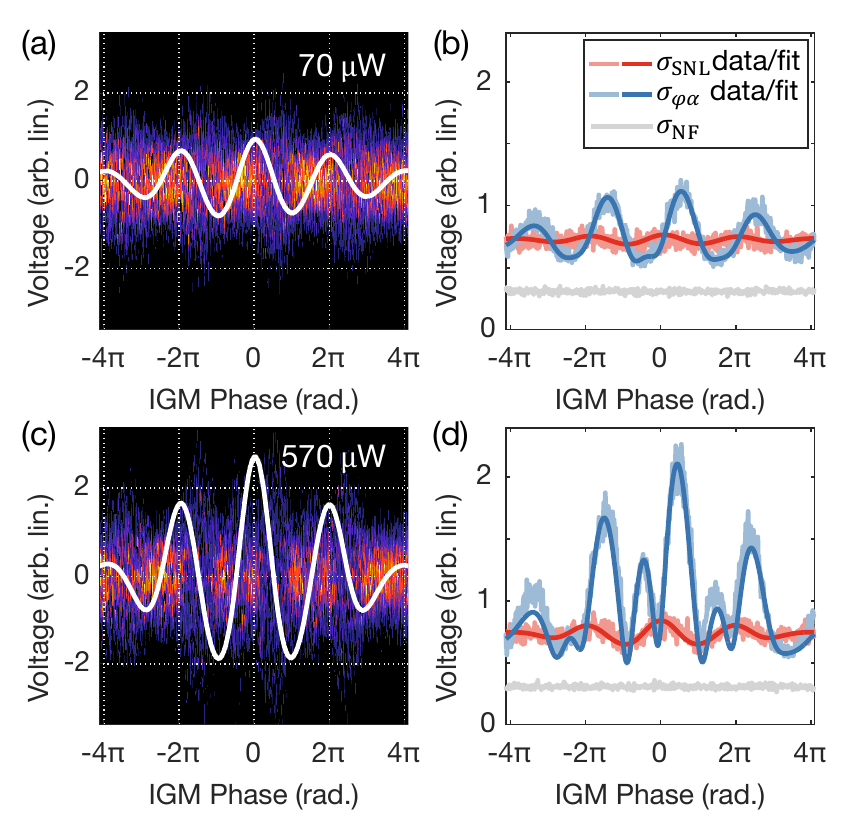}
\caption{Dual-detector measurements of quantum noise reduction in dual-comb centerbursts. (a) False-color histogram of noise-gram (sum) voltage counts for coherent power of 70 $\mu$W and squeezed power of 15 mW. White line is the average interferogram and brighter colors indicate higher voltage counts. (b) Corresponding voltage standard deviations for power ratio from (a). A sinusoidal non-stationary noise-gram pattern in the sum voltages (blue) is observed with one local maximum per 2$\pi$. The time-dependent SNL is calculated from the difference voltage (red). Fits to the data are shown as dark lines. The noisefloor of the measurement is shown in gray. (c) Same as (a) for coherent power of 570 $\mu$W and squeezed power of 15 mW. (d) Corresponding voltage standard deviations for power ratio from (c). A non-stationary pattern is observed in the sum voltages (blue) is observed with two local maxima per 2$\pi$. A larger variation of the SNL is also measured.}
\label{fig:5}
\end{figure}
The dual detector measurements are shown in Figs. 5(a)-(d). We clearly see the frequency doubling associated with the bifurcation.\ We notice that the squeezing is reduced after the bifurcation point although the anti-squeezing dramatically increases to $>$9 dB (see Supplemental Material). The highly non-classical behavior of the noise is nearly completely captured by our simple model given by Eq.\ (2). The model deviates from the data as the mode mismatch between the pulses increases, possibly due to the effects of temporal chirp and spectral mismatch. However, this effect may also arise from the distribution of CEO phases of our down-selected IGMs.\\
\indent This demonstration of phase-dependent quantum-noise in a dual-comb IGM is a critical step towards the development of quantum-enhanced dual-comb timing systems. With time-programmable combs \cite{Caldwell22,Caldwell23}, we can control the phase difference between the squeezed comb and coherent state comb such that the coherent state always displaces the squeezed comb to reveal sub-SNL behavior on the IGM fringe \cite{Herman18}. When a coherent state interferes with a squeezed comb, less optical power is required to achieve the same SNR on the time-programmable IGM fringe signal. This property enables low-power optical timing systems without the need for amplification, which could introduce additional noise and system complexity. For most timing and ranging applications, a balanced homodyne measurement setup is required \cite{Wang18}. Quantum advantage in balanced homodyne measurement will be achieved in the future through the use of a squeezed comb as a sensing probe and a large coherent state as a local oscillator (LO) \cite{Rubin07}.\\
\indent Our measurements also constitute a new method for continuous variable quantum state tomography \cite{Lvovsky09}. Although pulsed squeezing is a well-established technique, the exact nature of quantum states generated when ultra-short pulses interact with nonlinear media is not well understood \cite{Riek17,Stammer24}. For example, the degree of non-Gaussianity generated in such states is of significant relevance for quantum sensing and computing protocols \cite{Yanagimoto22}. DCI allows high-speed recovery of the Wigner plane representation for pulsed quantum states, including squeezed vacuum combs \cite{Yu01}. In a dual-comb experiment, many complete scans of the relative pulse phase are measured quickly, without the need for interferometric optical path length stabilization. Currently, our small displacement pulse performs only partial tomography of the squeezed pulse. Balanced homodyne measurements with a strong, spectrally-shaped LO comb will enable more complete tomographic characterization of multi-mode, pulsed squeezed states \cite{Cai17}. Dual-comb quantum tomography can also utilize spectral aliases of the IGM at frequencies above $f_\mathrm{rep}$ to enable full phase/amplitude reconstruction of the optical pulses \cite{Walsh23, Klee13}. With these improvements, DCI can extract valuable information about squeezed states generated by intense few-cycle pulses in a variety of media (e.g. $\chi^{(2)}$ solids, $\chi^{(3)}$ solids, atomic gases) \cite{Eto08, Gundogdu23, Stammer24}.\\
\indent In summary, we introduced DCI as a valuable tool for quantum metrology, using the technique to characterize the phase dependence of pulsed squeezing with two independent frequency comb lasers.
With this advance, we establish the ability to arbitrarily scan the relative phase between a quantum comb and a coherent state comb. Replacing differential optical path changes generated by piezo-electric transduction or electro-optic modulation with dual-comb scanning allows for high-speed time-programmable quantum measurement with long, arbitrary phase delays. Quantum DCI should lead to improvements for comb-based timing, ranging, and networking. DCI also provides a promising method for characterizing exotic quantum states generated via extreme nonlinear optical interactions.\\ \\ 
\indent \textit{Acknowledgements} -- The authors thank Shawn Geller for useful discussions. This research was supported by the Office of Naval Research (ONR N00014-22-1-2438), the National Science Foundation (NSF QLCI
Award OMA-2016244), the Natural Sciences and Engineering Research Council of Canada and Photonique Quantique Québec. MKK and MH acknowledge support from the NSF Graduate Research Fellowship Program (GRFP).\\ \\
\indent \textit{Data availability} -- The data that support the plots within
this Letter and other findings of this study are available
from the corresponding author upon reasonable request.


%

\section{Supplementary Materials}
\section{Supplementary Note 1: Quantum theory of dual-comb displacements and power scaling of cyclo-stationary quantum noise}
We begin by applying a simplified model to the quantum state of light at the output of the nonlinear optical fiber. Given some coherent state $|\beta\rangle$ entering the nonlinear fiber, we make a Gaussian approximation informed by the properties of the $\chi^{(3)}$ nonlinearity that is assumed to be of the form $(\hat{b}^\dagger\hat{b})^2$, where $\hat{b}$ is the annihilation operator for the coherent field. This Gaussian approximation is as follows:
\begin{equation}
    e^{ig(\hat{b}^\dagger\hat{b})^2}|\beta\rangle \approx |\beta\rangle_{r,\phi}
\end{equation}
where $|\beta\rangle_{r,\phi}$ denotes a squeezed and then the displaced state written as:
\begin{equation}
|\beta\rangle_{r,\phi} = D(\beta)S(r,\phi)|0\rangle,
\end{equation}
for displacement operator $D(\beta) = \exp\left(\beta\hat{b}^\dagger - \beta^*\hat{b}\right)$ and single mode squeezing operator $S(r,\phi) = \exp\left(\frac{r}{2}(e^{-2i\phi}\hat{b}^2 - e^{2i\phi}\hat{b}^{\dagger 2})\right)$. We let $g$ represent the nonlinearity accumulated in the fiber. This approximation is only valid for small values of $g$, meaning either small nonlinear cofficients or short fiber lengths. This approximation is chosen for several reasons and comes with several stipulations. Firstly, the Kerr effect does not change the photon-number statistics of the input field. This guarantees that the value of the mean field, $\beta$, does not change. More precisely, our approximation changes the mean of the field by an additional quantity of $\sinh^2(r)$, but this change is assumed to be much smaller than the mean field and can be ignored. In order to also leave the amplitude quadrature variance of the squeezed state unchanged in our approximation, the Kerr effect must deterministically generate a specific squeezing angle. Direct computation shows that the appropriate angle is:
\begin{equation}
    \phi = \pm \frac{1}{2}\cos^{-1}\left(\frac{\tanh(r)(2|\beta|^2+\cosh(2r)+1)}{2|\beta|^2}\right) - \theta
\end{equation}
where $\theta$ is the argument of $\beta = |\beta|e^{i\theta}$. Since the amount of squeezing applied by the Kerr effect depends on the mean field driving the nonlinearity we can further use the approximation suggested by Haus \cite{haus2000} to estimate the amount of squeezing:
\begin{equation}
    r \approx -\frac{1}{2}\log \left(1+2g^2|\beta|^4 - 2g|\beta|^2\sqrt{1+g^2|\beta|^4}\right).
\end{equation}
For our purposes we chose to directly fit $r$, using this equation only to infer the constant $g$\\
\indent Next, we consider the effects of the displacement field. Our Kerr-squeezed state is mixed on a mostly transmissive beamsplitter with another coherent field to perform a displacement. The displacement field is allowed to populate a different temporal mode than the squeezed field. To model this operationally we introduce mode labels $\xi_1$ for the Kerr-squeezed state and $\xi_2$ for the displacement field. For simplicity, we assume that we implement an ideal displacement $D(\alpha,\xi_2)$. Under this displacement, the field becomes:
\begin{equation}
    D(\alpha,\xi_2)|\beta,\xi_1\rangle_{r,\phi} = |\beta+\gamma\alpha\rangle_{r,\phi}\otimes|\sqrt{1-|\gamma|^2}\alpha,\xi_\perp\rangle,
\end{equation}
where $\gamma$ is the complex mode overlap defined as $\gamma = \int dt \xi_1^*(t)\xi_2(t)$, and the perpendicular mode $\xi_\perp$ is defined as \cite{Lordi24}: 
\begin{equation}
    \xi_\perp = \frac{\xi_2 - \gamma \xi_1}{\sqrt{1-|\gamma|^2}}.
\end{equation}
The mode overlap parameter can be determined from the time-dependent function that describes the envelope and phase of the dual-comb interferogram (IGM) with the caveat that our modes $\xi_1$ and $\xi_2$ are normalized, making $|\gamma|^2 \le 1$.\\
\indent Now that we have a model for the quantum field, we need a model for the photo-detection. We assume that our detector directly measures the photon number operator represented by $\hat{b}^\dagger\hat{b}$ and that this detector has an effective efficiency $\eta$. We further assume that the detector has a slow response and cannot resolve the temporal profile of the pulses from either comb, and each detector bin fully contains one pulse from each comb. This approximation is valid for a detection bandwidth that is comparable to the repetition rate of the combs, and thus the measurement bin is much longer than the comb pulse width.\\
\indent Because we take a Gaussian approximation for the field, we can completely describe the measurement statistics through an analysis of the first two moments. First, we model the inefficient detector with efficiency $\eta$ as an ideal detector preceded by a virtual beamsplitter with transmissivity $\eta$. This loss operation will reduce the incoming signal, but will also introduce additional vacuum fluctuations. To understand the effect of efficiency we will modify the mean and variance of our signal according to the effective loss. This operation reduces the mean of the measurement by a factor of $|\eta|$, that is:
\begin{equation}
    \langle \hat{b}^\dagger\hat{b}\rangle \to |\eta|\langle\hat{b}^\dagger\hat{b}\rangle.
\end{equation}
 While the variance also decreases by a factor of $|\eta|^2$, the vacuum port of the virtual beamsplitter also contributes another term $|\eta(1-\eta)|\hat{b}^\dagger\hat{b}$. This term persists after taking the vacuum expectation value and comes from the normal ordering of the creation and annihilation operators in the vacuum mode. The variance of direct detection with inefficient detection is thus:
\begin{equation}
    \langle (\Delta \hat{b}^\dagger\hat{b})^2\rangle \to |\eta|^2\langle (\Delta \hat{b}^\dagger\hat{b})^2\rangle +|\eta(1-\eta)|\langle\hat{b}^\dagger\hat{b}\rangle,
\end{equation}
where $\langle (\Delta \hat{b}^\dagger\hat{b})^2\rangle = \langle ( \hat{b}^\dagger\hat{b})^2\rangle -\langle\hat{b}^\dagger\hat{b}\rangle^2$.\\
\indent We now calculate the first and second moments of the photon number operator for our displaced Kerr-squeezed state. The Gaussian approximation demands that the first two moments define the distribution and we conveniently determine the means and variances of $\xi_1$ and $\xi_\perp$ separately, and then add them according to the properties of normal distributions. The perpendicular mode measurement distribution is the easiest to calculate since it is in a coherent state:
\begin{equation}
\begin{aligned}
     \langle\sqrt{1-|\gamma|^2}\alpha|\hat{b}^\dagger\hat{b}|\sqrt{1-|\gamma|^2}\alpha\rangle &= (1-|\gamma|^2)|\alpha|^2\\
     \langle\sqrt{1-|\gamma|^2}\alpha|(\Delta \hat{b}^\dagger\hat{b})^2|\sqrt{1-|\gamma|^2}\alpha\rangle &= (1-|\gamma|^2)|\alpha|^2.
\end{aligned}
\end{equation}\\
\indent For the squeezed mode, we apply a Mollow transform to move our coordinates to the frame defined by the mean field $\beta+\gamma\alpha$. For convenience, we write $\beta+\gamma\alpha = \nu = |\nu|e^{i\varphi}$ and then expectation value follows as:
\begin{equation}
\begin{aligned}
    \!_{r,\phi}\langle\nu|\hat{b}^\dagger\hat{b}|\nu\rangle_{r,\phi} &= \!_{r,\phi}\langle 0|(\hat{b}^\dagger+\nu^*)(\hat{b}+\nu)|0\rangle_{r,\phi}\\
    &=\!_{r,\phi}\langle 0|\hat{b}^\dagger\hat{b} + |\nu|(e^{i\varphi}\hat{b}^\dagger + e^{-i\varphi}\hat{b}) + |\nu|^2|0\rangle_{r,\phi}.
\end{aligned}
\end{equation}\\
\indent We notice $|\nu|(e^{i\varphi}\hat{b}^\dagger + e^{-i\varphi}\hat{b}) = \sqrt{2}|\nu|\hat{x}_\varphi$, where $\hat{x}_\varphi$ is the quadrature operator $\hat{x}_\varphi = \cos(\varphi)\hat{x} + \sin(\varphi)\hat{p}$ for position and momentum operators $\hat{x}$ and $\hat{p}$ respectively. Squeezed vacuum has a mean of 0 in any quadrature and a photon number mean of $\sinh^2(r)$ so the mean reduces to:
\begin{equation}
\!_{r,\phi}\langle\nu|\hat{b}^\dagger\hat{b}|\nu\rangle_{r,\phi} = |\nu|^2 +\sinh^2(r).
\end{equation}
The mean field term, $|\nu|^2$, is of order $10^8$ at the specific power and wavelength of the lasers used here. Even for the most optimistic amounts of squeezing $\sinh^2(r)$ will be of order $10^2$, meaning that this term is negligible.\\
\indent The measurement variance requires us to find the second moment of the photon number distribution:
\begin{equation}
    \begin{aligned}
        \!_{r,\phi}\langle\nu|(\hat{b}^\dagger\hat{b})^2|\nu\rangle_{r,\phi} &= \!_{r,\phi}\langle 0|\left(\hat{b}^\dagger\hat{b} + \sqrt{2}|\nu|\hat{x}_\varphi + |\nu|^2\right)^2|0\rangle_{r,\phi}\\
        &= \!_{r,\phi}\langle 0|(\hat{b}^\dagger\hat{b})^2+ 2|\nu|^2\hat{x}_\varphi^2 + |\nu|^4 + \sqrt{2}|\nu|\hat{b}^\dagger\hat{b}\hat{x}_\varphi +\sqrt{2}|\nu|\hat{x}_\varphi\hat{b}^\dagger\hat{b} \\&+ 2|\nu|^2\left(\hat{b}^\dagger\hat{b}+\sqrt{2}|\nu|\hat{x}_\varphi\right) |0\rangle_{r,\phi}.
    \end{aligned}
\end{equation}
Now go term by term through this expectation value as follows:
\begin{equation}
    \begin{aligned}
        \!_{r,\phi}\langle 0|(\hat{b}^\dagger\hat{b})^2|0\rangle_{r,\phi} &= 3\sinh^4(r) +2\sinh^2(r)\\
         \!_{r,\phi}\langle 0|2|\nu|^2\hat{x}_\varphi^2|0\rangle_{r,\phi} &= |\nu|^2\left(\cosh(2r)-\sinh(2r)\cos(2(\phi-\varphi))\right)\\
         \!_{r,\phi}\langle 0||\nu|^4|0\rangle_{r,\phi} &= |\nu|^4\\
         \!_{r,\phi}\langle 0|2|\nu|^2\left(\hat{b}^\dagger\hat{b}+\sqrt{2}|\nu|\hat{x}_\varphi\right)|0\rangle_{r,\phi} &= 2|\nu|^2\sinh^2(r)\\
         \!_{r,\phi}\langle 0|\sqrt{2}|\nu|\hat{b}^\dagger\hat{b}\hat{x}_\varphi|0\rangle_{r,\phi}   &= \!_{r,\phi}\langle 0|\sqrt{2}|\nu|\hat{x}_\varphi\hat{b}^\dagger\hat{b}|0\rangle_{r,\phi} =  0.
    \end{aligned}
\end{equation}
The first four of these terms are standard in Gaussian quantum optics and can be found in \cite{Caves2025}. The final terms are less common but are calculated quickly by considering that $\hat{b}^\dagger\hat{b}\hat{x}_\varphi$ can be split into two terms, one proportional to $\hat{b}^\dagger\hat{b}^2$ and another to $\hat{b}^\dagger\hat{b}\hat{b}^\dagger$. Both of these terms have an odd total number of creation and annihilation operators and squeezed vacuum only has support on even Fock states, so this expectation must be zero.\\
\indent Using this information and subtracting the mean squared, we get the photon number variance:
\begin{equation}
    \!_{r,\phi}\langle\nu|(\Delta \hat{b}^\dagger\hat{b})^2|\nu\rangle_{r,\phi} = 2\sinh^4(r)+2\sinh^2(r) + |\nu|^2\left(\cosh(2r)-\sinh(2r)\cos(2(\phi-\varphi))\right).
\end{equation}
To get the final variance of this measurement we add on the mean and variance from the perpendicular mode and take into account the efficiency to get the modeled mean and variance: 
\begin{equation}
    \begin{aligned}
        \text{Mean} &= |\eta|\left(|\beta+\gamma\alpha|^2 +\sinh^2(r) + (1-|\gamma|^2)|\alpha|^2\right)\\
        \text{Variance} &= |\eta|^2\left[ 2\sinh^4(r)+2\sinh^2(r) + |\beta+\gamma\alpha|^2\left(\cosh(2r)-\sinh(2r)\cos(2(\phi-\varphi))\right)\right]\\ &+ |\eta(1-\eta)|\left[|\beta+\gamma\alpha|^2 +\sinh^2(r) \right] + |\eta|(1-|\gamma|^2)|\alpha|^2
    \end{aligned}
\end{equation}
We then group the terms according to their physical origin:
\begin{equation}
    \begin{aligned}
        \text{Mean} &= |\eta|\left(\underbrace{|\beta|^2 +|\alpha|^2}_{\text{Total Power}} + \underbrace{2\text{Re}(\beta^*\gamma\alpha)}_{\text{IGM}} +\underbrace{\sinh^2(r)}_{\text{Squeezed Power}}\right)\\
        \text{Variance} &= |\eta|^2\left[\underbrace{ 2\sinh^4(r)+ \sinh^2(r)}_{\text{Squeezed Number Variance}} + \underbrace{|\beta+\gamma\alpha|^2\left(\cosh(2r)-\sinh(2r)\cos(2(\phi-\varphi)) - 1\right)}_{\text{Squeezed Amplitude Variance}}\right]\\ &+ |\eta|\left(\underbrace{|\beta|^2 +|\alpha|^2}_{\text{Total Shot Noise}} + \underbrace{2\text{Re}(\beta^*\gamma\alpha)}_{\text{IGM Shot Noise}} +\underbrace{\sinh^2(r)}_{\text{Squeezed Shot Noise}}\right).
    \end{aligned}
\end{equation}
We also make the approximation that the mean field is large with respect to the squeezing power, $|\beta|^2+|\alpha|^2 \gg \sinh^2(r)$, giving the simpler form:
\begin{equation}
    \begin{aligned}
        \text{Mean} &= |\eta|\left(|\beta|^2 +|\alpha|^2 + 2\text{Re}(\beta^*\gamma\alpha)\right)\\
        \text{Variance} &= |\eta|^2|\beta+\gamma\alpha|^2\left(2\sinh^2(r)-\sinh(2r)\cos(2(\phi-\varphi))\right)\\
        &+ |\eta|\left(|\beta|^2 +|\alpha|^2 + 2\text{Re}(\beta^*\gamma\alpha)\right).
    \end{aligned}
\end{equation}\\
\indent This expression can also be written in terms of the IGM phase, or equivalently the phase of state $\alpha$. We define $\varphi_\beta$ and $\varphi_\alpha$ as the argument of $\beta$ and $\alpha$, respectively. Simple geometry analysis reveals that: 
\begin{equation}
    \varphi = \arctan\left(\frac{|\beta|\sin(\varphi_\beta)+|\gamma\alpha|\sin(\varphi_\alpha)}{|\beta|\cos(\varphi_\beta)+|\gamma\alpha|\cos(\varphi_\alpha)}\right).
\end{equation}
We simplify this expression by fixing $\beta$ as a real number (arbitrary choice of reference phase) and then assuming $|\beta|\gg|\alpha|$,
\begin{equation}
        \varphi \approx \arctan\left(\frac{|\gamma\alpha|\sin(\varphi_\alpha)}{|\beta|}\right)\approx \frac{|\gamma\alpha|\sin(\varphi_\alpha)}{|\beta|}.
\end{equation}

\nocite{*}
\begin{figure}[!h]
    \centering
    \includegraphics[width=0.8\linewidth]{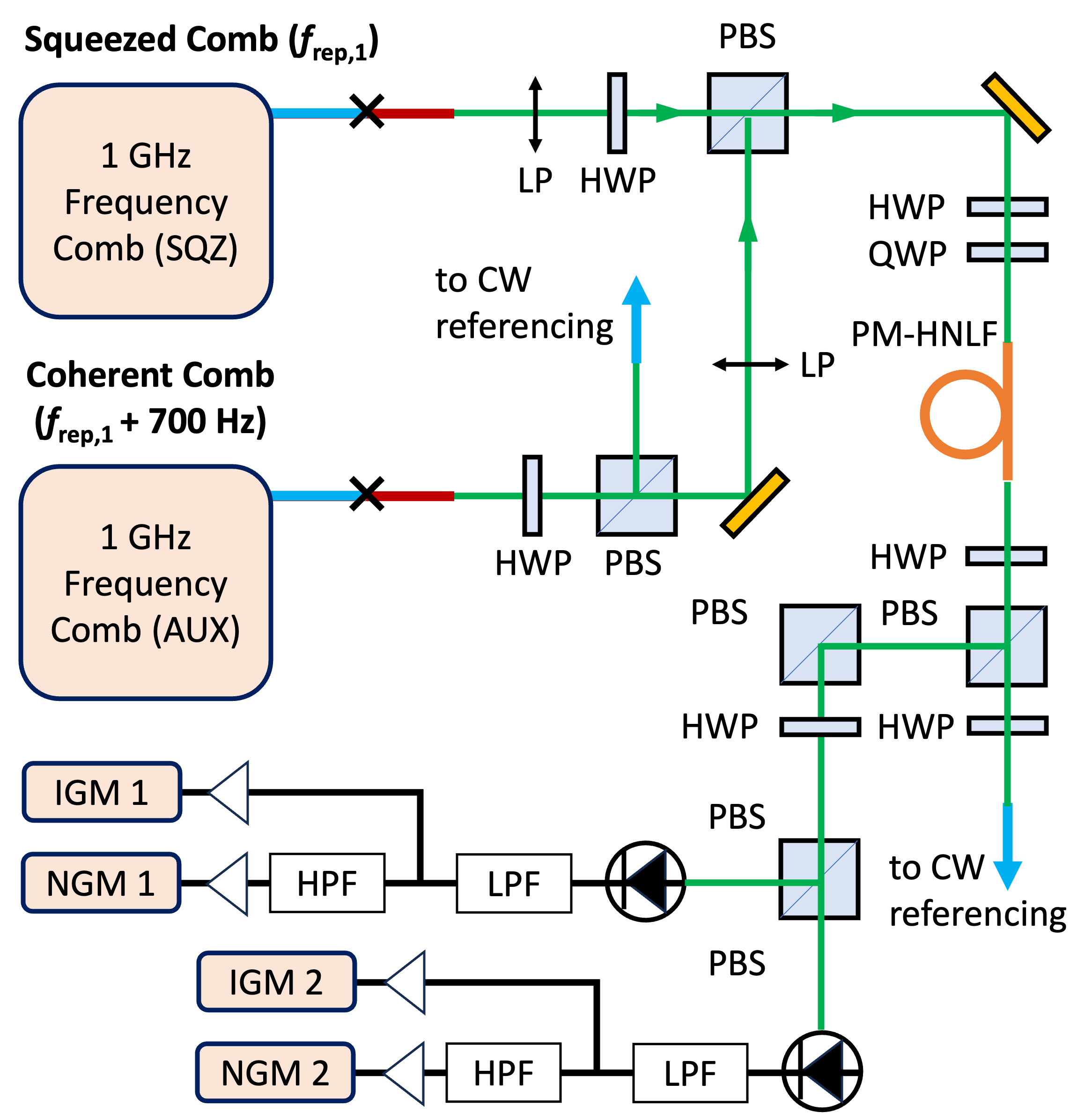}
    \caption{Schematic of two-detector experiment. The setup is nearly identical to the single-detector version except we place a half-wave plate (HWP) and a polarizing beamsplitter (PBS) to act as a 50:50 beamsplitter. We digitize the four-signals IGM1, IGM2, NGM1 and NGM2 using a high-speed four-channel analog-to-digital converter. LP = linear polarizer; QWP = quarter-wave plate; PM-HNLF = polarization-maintaining highly nonlinear fiber; LPF = low-pass RF filter; HPF = high-pass RF filter; Blue lines indicate PM-1550 fiber; Red lines indicate PM normal dispersion fiber; Green lines indicate free space beams; Orange line indicates PM HNLF.}
    \label{fig:s1}
\end{figure}
\begin{figure}[!h]
    \centering
    \includegraphics[width=0.8\linewidth]{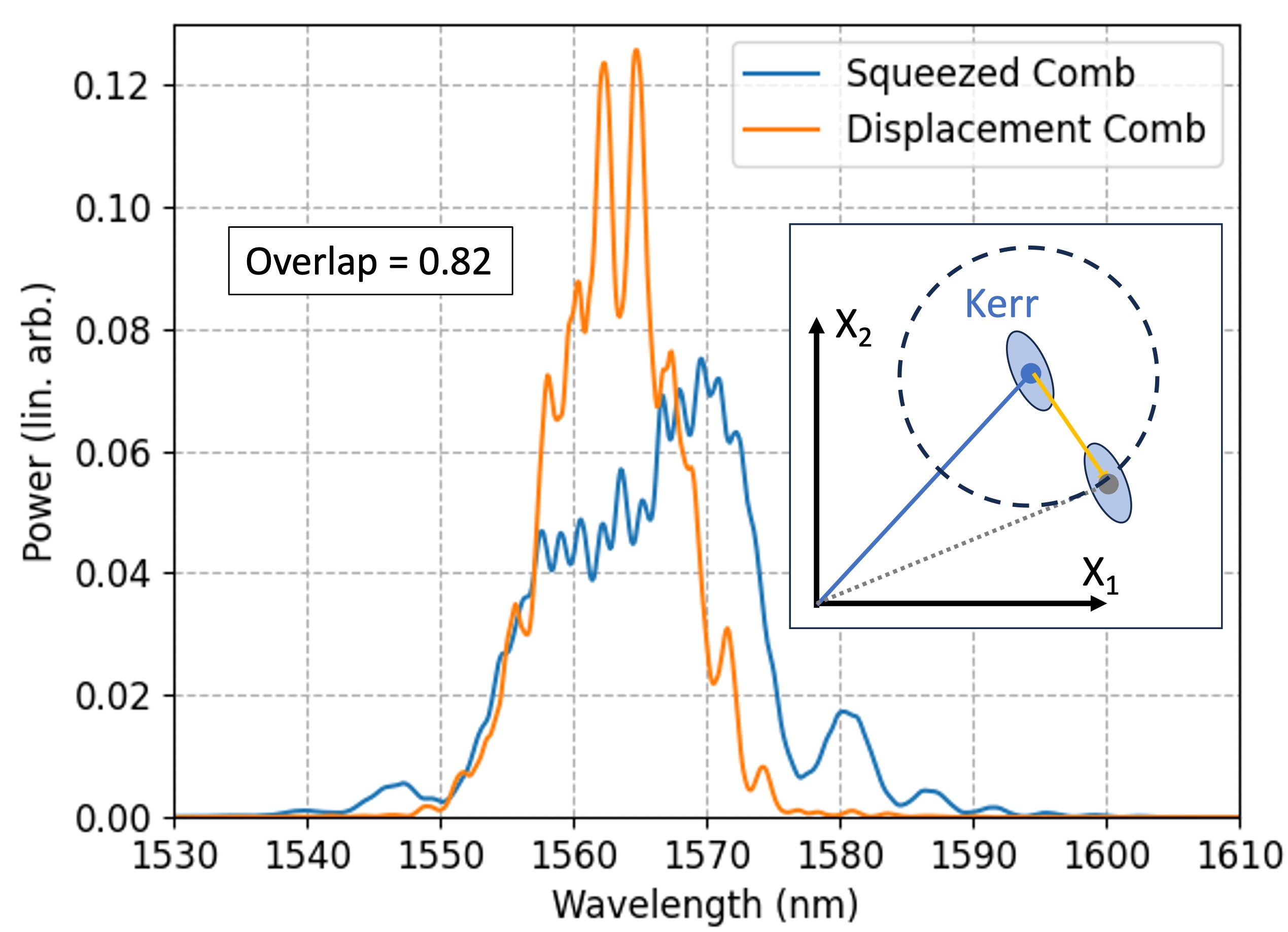}
    \caption{Comb spectra measured with an optical spectrum analyzer. Displacement comb (coherent state comb) is shown in orange and squeezed comb is shown in blue. Inset: Phase space schematic of displacement of bright Kerr squeezed state by a small coherent state.}
    \label{fig:s2}
\end{figure}
\begin{figure}[!h]
    \centering
    \includegraphics[width=1\linewidth]{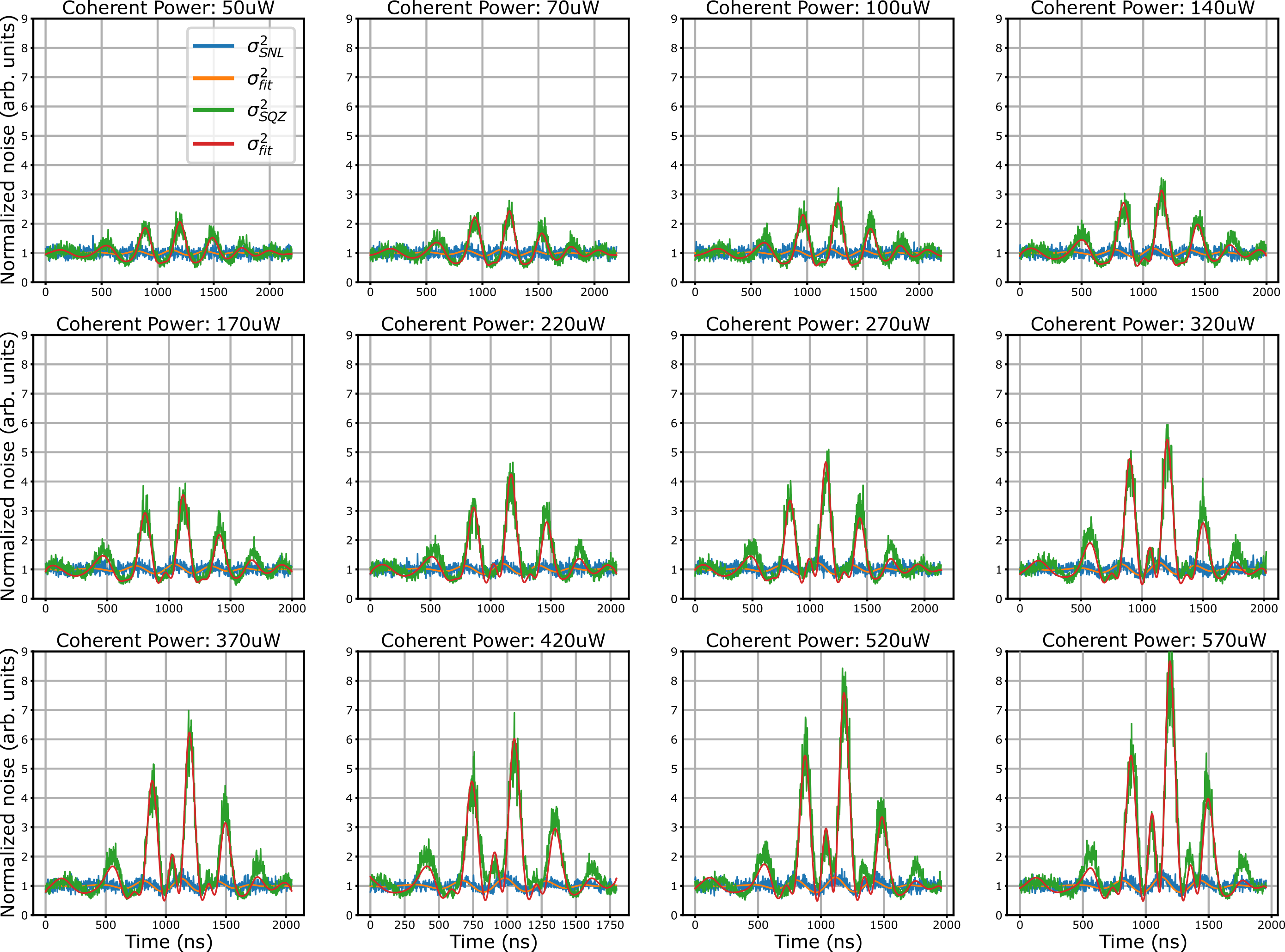}
    \caption{Results of fit to Eq. (19) for various coherent comb powers ranging from 50 $\mu$W to 570 $\mu$W. The model successfully captures the frequency doubling effect as the displacement pushes the Kerr state past the point of optimal amplitude squeezing ($\sim$150 $\mu$W). SNL = shot-noise limit (difference signal); SQZ = squeezed amplitude variance (sum signal).}
    \label{fig:s3}
\end{figure}

\begin{figure}[!h]
    \centering
    \includegraphics[width=1\linewidth]{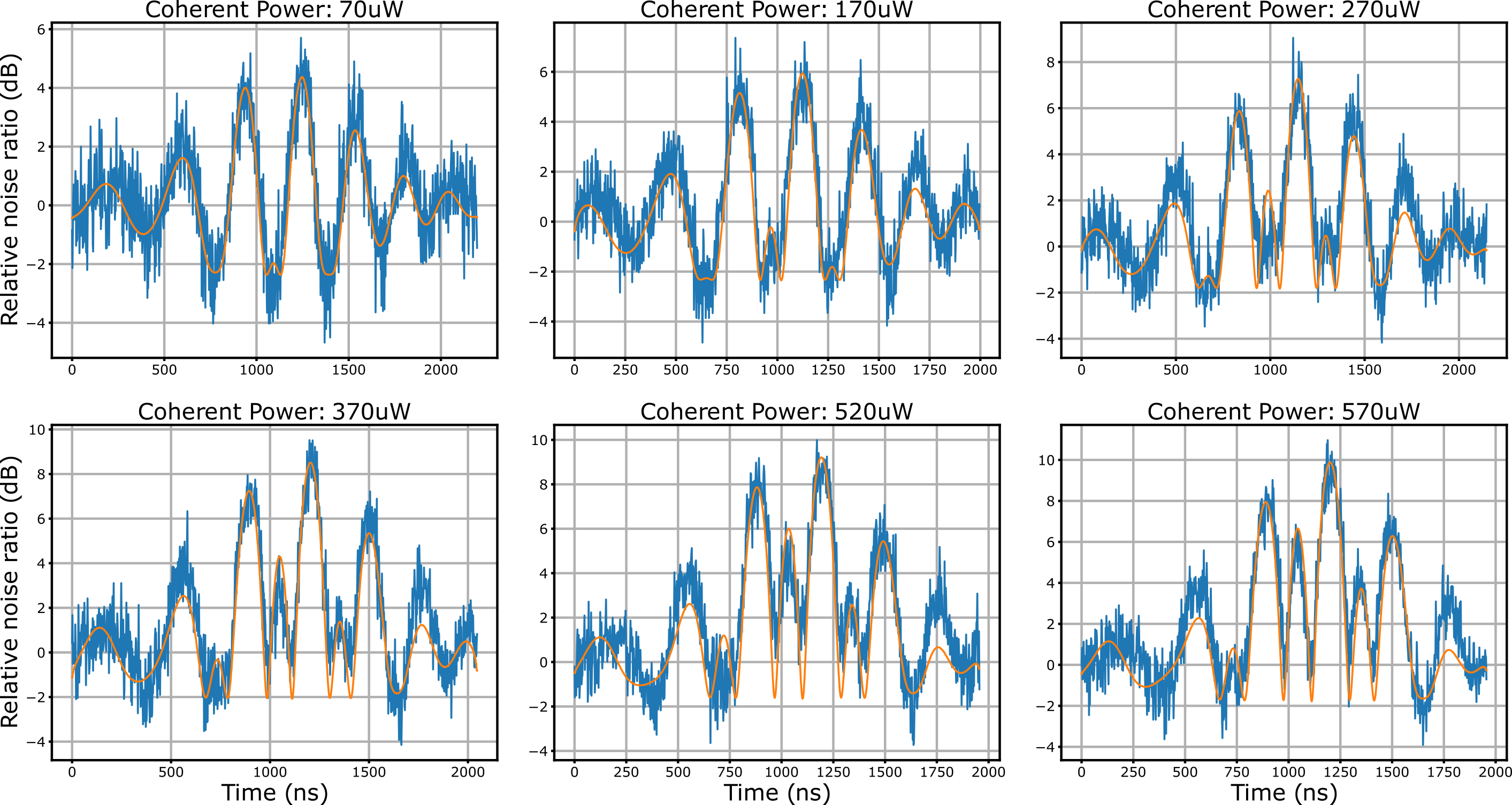}
    \caption{Results of fit to Eq. (19) expressed as squeezing ratio. Maximum squeezing of $\sim$3 dB is measured for coherent power of 170 $\mu$W. Anti-squeezing continues to grow as displacement strength increases.}
    \label{fig:s4}
\end{figure}
\section{Supplementary Note 2: Complementary measurements of noise-grams using two detector setup}
As discussed in the main text, we implemented a two-detector experiment to verify the presence of squeezing in the dual-comb centerburst. A schematic of the setup is shown in Fig. 6. Following previous demonstrations of amplitude squeezing \cite{Paschotta94,Fiorentino01}, we use the difference of the two detectors as the shot noise limit, and the sum of the two detectors as the amplitude-squeezed signal. Unlike other experiments, we perform the difference/sum operations in post-processing after digitization of the individual photodetector outputs. The data is taken for a fixed squeezed power of $\sim$14.7 mW and coherent displacement power between 50 $\mu$W and 570 $\mu$W, taking care to avoid photodetector saturation and overfilling the analog-to-digital converter's dynamic range. Like the single detector experiments, the IGM carrier frequency is set to 3 MHz and the repetition rate difference is set to $\sim$700 Hz. The optical spectra for the squeezed comb and displacement combs are shown in Fig. 7. Note that the normalized spectral overlap between the two laser combs is 0.82. We cross-correlate the IGMs from each beamsplitter output to confirm that there is no measurable delay between the two digitized channels.\\
\indent Numerical fits to the model from Eq. (19) at varying coherent displacement comb power levels are shown in Fig. 8. The fit utilizes the IGM amplitudes and phases recovered using a Hilbert transform. The results of Hilbert transform also also used to down-select the IGMs with similar IGM CEO phase (difference between comb 1 CEO phase and comb 2 CEO phase). The down-selected IGMs have CEO phases within a range of approximately 50 milliradians. The mean field value of the displaced squeezed state is used to fit the shot noise level (difference signal) during the IGM. The amplitude variance from Eq. (19) is used to model the sum signal. There is a very close match between theory and experiment for the central fringes of the noise-grams (NGMs). This match becomes less clear towards the edges of NGMs, possibly due to the unmodeled effects of chirp or spectral mode mismatch but may also simply be an artifact of the slight variation in CEO phase between the downselected IGMs. Future experiments with completely self-referenced combs will help to verify the limits of our simple model. Although the Gaussian model recreates most of the features seen in the measurement data, we should not consider this model to be physically complete. We expect the $\chi^{(3)}$ nonlinearity to generate a non-Gaussian state \cite{haus2000}. In the limit of small squeezing and due to the combined effects of loss and phase noise, we produce a state that is well-approximated by a Gaussian state. Since our experiment does not perform a tomographically complete set of measurements we are not able to conclusively determine the state leaving the fiber. We also note that large differences between the noise of the squeezed and antisqueezed quadrature (see Fig. 9) drives the fit algorithm to find an efficiency parameter ($\eta$) between $0.3$ and $0.4$, which is much lower than the estimated loss after the fiber of about $75\%$. In the model, this efficiency is a loss prior to detection, but it is possible that another physical mechanism is increasing the product of position and momentum uncertainty (e.g., Raman or Brillouin scattering in the nonlinear fiber \cite{Dong08}). It is possible that spectral mode mismatch and spatial mode mismatch of the two orthogonally-polarized modes of the highly nonlinear fiber lead to further degradation of the efficiency of the displacement operation. The current model successfully matches the observed squeezing in the two-detector experiment ($\sim 2\text{dB}$) but results in un-physically large values for the theoretical squeezing ($\sim28\ \text{dB}$) in the presence of no loss. This observation suggests that although the final state is well approximated by Gaussian quantum statistics, the simple model does not completely capture the quantum dynamics of the Kerr nonlinearity in the fiber \cite{White00}. This motivates future research into tomography of Kerr states and more complete non-Gaussian modeling of pulse dynamics in nonlinear media \cite{Yanagimoto22}.
\section{Supplementary Note 3: Future Improvements to Dual-Comb Centerburst Squeezing}
We now consider the implications of our results for quantum sensing. One potential use-case is optical networking with multiple nodes. In this scenario, one optical pulse train is used to synchronize multiple points in a measurement chain (\textit{e.g.} at a free-electron laser or synchrotron facility). Here, it may be useful to know the pulse train timing without fully depleting the signal since the pulses will be used for later comparisons and experiments. In this case, a displacement-like measurement as performed in our work might be useful since we can enhance the SNR on the measurement using the large squeezed state and still have signal power left over for subsequent use. As mentioned in the main text, optical timing experiments with squeezed combs can be applied more generally when a balanced homodyne experiment is performed using a small squeezed comb and a stronger local oscillator comb. Currently, our ability to use a strong local oscillator is limited by photodetector saturation and analog-to-digital converter dynamic range. For the purpose of noise analysis, our demonstrated electrical scheme (splitting the IGM and NGM signals) successfully circumvents ADC dynamic range limitations, but photodiode saturation still presently limits our ability to perform homodyne-like measurements of squeezed states using dual-comb interferometry. To mitigate saturation effects and expand the intensity range of squeezed states that can be characterized by linear DCI without separate NGM amplification, future experiments can incorporate multi-detector arrays \cite{Fiorentino01}. Another option is to generate squeezed states with small mean fields using balanced nonlinear interferometry schemes \cite{Yu01}. With fully stabilized combs, the carrier-envelope IGM phase can be set to zero which will allow for every IGM to be analyzed rather than down-selecting in post-analysis.
\\

\indent 


%

\bibliography{DIH}
\end{document}